\documentclass[aps,prl,twocolumn,showpacs,superscriptaddress,groupedaddress]{revtex4}  % for review and submission
\usepackage{graphicx,tensor, cleveref}  % needed for figures
\usepackage{dcolumn}   % needed for some tables
\usepackage{bm}        % for math
\usepackage{amssymb}   % for math
\usepackage{amsmath}
\usepackage{dsfont}
\usepackage{cancel}
\usepackage{xcolor}

\begin{document}

% The following information is for internal review, please remove them for submission
%*\widetext
%*\leftline{Version xx as of \today}
%*\leftline{Primary authors: Joe E. Physics}
%*\leftline{To be submitted to (PRL, PRD-RC, PRD, PLB; choose one.)}
%*\leftline{Comment to {\tt d0-run2eb-nnn@fnal.gov} by xxx, yyy}
%*\centerline{\em D\O\ INTERNAL DOCUMENT -- NOT FOR PUBLIC DISTRIBUTION}

% the following line is for submission, including submission to the arXiv!!
%\hspace{5.2in} \mbox{Fermilab-Pub-04/xxx-E}

\title{Lorentzian geometry of qubit entanglement}
\author{ Joseph Samuel, Kumar Shivam and  Supurna Sinha}
%\address{Raman Research Institute, Bangalore 560080, India.}
%*\input author_list.tex       % D0 authors (remove the first 3 lines
                             % of this file prior to submission, they
                             % contain a time stamp for the authorlist)
                             % (includes institutions and visitors)

%\author{Joseph Samuel$^{1}$, Kumar Shivam$^{1}$\ and
%Supurna Sinha$^{}$}
\address{Raman Research Institute, Bangalore 560080, India.}

\date{\today}

\begin{abstract}
We study the relation between qubit entanglement and Lorentzian
geometry. In an earlier paper, we had given a recipe 
for detecting two qubit entanglement. The entanglement criterion is  
based on Partial Lorentz Transformations (PLT) on individual qubits. 
The present paper gives the theoretical framework underlying 
the PLT test.
The treatment is based physically, on the causal structure of Minkowski
spacetime, and mathematically, on a Lorentzian 
Singular Value Decomposition. A surprising feature is the 
natural emergence of ``Energy conditions'' used in Relativity.
All states satisfy a ``Dominant Energy Condition'' 
(DEC) and separable states satisfy the 
Strong Energy Condition(SEC), while entangled states violate the SEC. 
Apart from testing for entanglement, our approach also 
enables us to construct a separable form for the 
density matrix in those cases where it exists. 
Our approach leads to a simple graphical three 
dimensional representation of the state space 
which shows the entangled states within the set of all states.

\end{abstract}

\maketitle

%\section{\label{sec:level1}First-level heading}
% sections are not used for PRL papers

%-------------------------------------------------------------

\section{I. Introduction}

Detecting entanglement is one of the outstanding problems
in Quantum Information Theory.
In two qubit systems, the Positive Partial Transpose (PPT) criterion 
\cite{PhysRevLett.77.1413,horodeckipla,geometry} 
gives a simple, computable criterion for detecting entanglement. 
The criterion gives a necessary and suficient condition for a 
state to be separable.

In an earlier paper\cite{earlier}, we 
proposed a new test based on 
Partial Lorentz Transformation(PLT) of 
individual qubits. It turns out that like the PPT test, 
the PLT criterion is necessary and sufficient in the two qubit case.  
In \cite{earlier}, the PLT test was given as a recipe, a form that could be
directly used by those who want to apply the test.  
The purpose of the present paper is to describe the theoretical framework
behind the PLT test. 
In addition to showing why the test works, 
our Lorentzian approach yields an explicit separable form 
of the density matrix, when such a form exists. 
It also permits a complete elucidation
of the state space using a Lorentzian version of the 
Singular Value Decomposition. 
The PLT test uses ideas borrowed from the space-time physics of 
Special Relativity.

The paper is organized as follows. In Section II we discuss 
Partial Lorentz Transformations (PLT). 
Section III describes 
the Lorentzian Singular Value Decomposition
which provides the theoretical basis for the PLT test.
Section IV gives necessary and sufficient conditions on the singular
values to define a state and expresses the state in separable form,
under certain conditions on the singular values. We also show that these conditions
are necessary for separability. 
We then discuss a simple three  dimensional representation of 
the two-qubit state space in Section V. Section VI deals with 
non generic states. 
We finally end the paper with some concluding remarks in Section VII.

We use a Lorentzian metric of signature mostly minus: $g=diag(1,-1,-1,-1)$. Spacetime 
Lorentz indices $\mu,\nu$ range over $0,1,2,3$, as also do Frame indices $a,b,..$. Both these indices
are raised and lowered by the Minkowski metric and we use the Einstein summation convention.  
All causal (timelike or lightlike 4-vectors)
are pointing into the future. 
Throughout this paper, by ``Lorentz group'', we mean 
its proper, orthochronous subgroup, which preserves time orientation
as well as the spatial orientation.

%-------------------------------------------------------------

\section{II. Lorentz Transformations}

The states of a qubit can be expressed in space-time 
form by using $\sigma_{\mu}=(\mathds{1}, \sigma_x, \sigma_y, \sigma_z)$, 
the identity and the Pauli matrices
\begin{equation}\label{1qubit}
\tau=u^{\mu}\sigma_{\mu}
\end{equation}
$u^{\mu}$ is a real future pointing $4$-vector and satisfies 
\begin{equation}
u^{\mu}u^{\nu}g_{\mu\nu}>0
\end{equation}
for impure states and 
\begin{equation}
u^{\mu}u^{\nu}g_{\mu\nu}=0
\end{equation}
for pure states. 
Impure states have time-like $u$ and pure states have lightlike $u$. 
In both the cases $u^0>0$, the 4-vector $u^{\mu}$ is future pointing. 
If we were to fix the ``normalization" by $\textrm{Tr}(\rho)=2$, 
$u^{0}=1$, 
the impure states can be represented in the Bloch ball $\vec{u}.\vec{u}<1$ and the 
pure states on the Bloch sphere $\vec{u}.\vec{u}=1$. The Lorentzian nature of
the state space is already evident.
Under Lorentz Transformations
\[u\indices{^{\mu}}\mapsto u\indices{^{\prime\mu}}=S\indices{^{\mu}_{\nu}}u\indices{^{\nu}} \]
where $S\indices{^{\mu}_{\nu}}S\indices{^{\alpha}_{\beta}}g\indices{_{\mu\alpha}}=g\indices{_{\nu\beta}}$. 
The Lorentz Transformation maps states to states. 
The group action has two orbits: the pure states constitute one orbit
and the impure states another.

{\it Partial Lorentz Transformations:}
Let $\rho$ be a density matrix of a two qubit system. We 
assume $\rho$ is non negative ($\rho\ge 0$), 
Hermitian ($\rho^{\dagger}=\rho$).
In our treatment, we will not need to normalize $\rho$,
but we suppose $\rho$ does not vanish identically.
One can expand the density matrix $\rho$ as
\begin{equation}
        \rho=\frac{1}{4} A^{\mu\nu}\sigma_{\mu}\otimes\sigma_{\nu}
\end{equation}
where $A^{\mu\nu}$ can be calculated from
\begin{equation}\label{A}
 A_{\mu\nu}=\textrm{Tr}(\rho\sigma_{\mu}\otimes\sigma_{\nu}).
\end{equation}

Consider doing a 
Lorentz Transformation on just the first subsystem
\begin{equation}
\sigma\indices{_{\mu}}\mapsto \sigma\indices{_{\mu}^'}=\sigma\indices{_{\alpha}} L\indices{^{\alpha}_{\mu}}.
\end{equation}
This results in a new state $\rho^{\prime}=\frac{1}{4}L\indices{^{\mu}_{\alpha}}A\indices{^{\alpha\nu}}
\sigma\indices{_{\mu}}\otimes\sigma\indices{_{\nu}}$, so 
\[A\indices{^{\prime\mu\nu}}=L\indices{^{\mu}_{\alpha}}A\indices{^{\alpha\nu}}.\]
We refer to this as a Partial Lorentz Transformation since it acts only on the first subsystem. Similarly 
one can perform a Partial Lorentz Transformation on the second subsystem
\[ A\indices{^{\prime\prime\mu}^{\nu}} = A\indices{^{\mu}^{\alpha}}R\indices{^{\nu}_{\alpha}}. \]
Partial Lorentz Transformations act on $A$ by left (L) and right (R) actions. It is elementary to 
check that PLT s are completely positive\cite{geometry} 
maps on the state space. 
%expand?
They also have the 
important property that they 
preserve separability of states. The PLT of a separable state is separable. 
The PLT of an entangled state is entangled. 
This is the key property of the Partial Lorentz Transformation group 
that we exploit here. 

\section{III. Lorentzian Singular Value Decomposition}
Let us now consider the action of left and right PLTs on the state space. The space of 
(unnormalized) density matrices is 16 dimensional. The left and the right PLTs 
generate orbits which are generically $6+6=12$ dimensional. Thus the 16 dimensional 
state space splits into a 4 parameter family of 12 dimensional fibers.
 (There are also isolated points where the isotropy subgroup is larger and the fiber smaller). 
Each fiber is either entirely separable or entirely entangled. 
Thus we can reduce the problem to the 4 dimensional space of orbits. 
In order to characterize the orbits,  consider
\begin{equation}\label{B}
B\indices{^{\mu}_{\nu}}=A\indices{^{\mu}_{\alpha}}A\indices{_{\nu}^{\alpha}}.
\end{equation}
B is obviously symmetric $B\indices{^{\mu\nu}}=B\indices{^{\nu\mu}}$.
It is easily checked that $\textrm{Tr}(B^n)$ is 
invariant under both left and right PLTs. Generically we would
expect the four eigenvalues of 
$B\indices{^{\mu}_{\nu}}$ to characterise the orbits. 

Just as we constructed $B$ from a state $A$, we 
can also similarly define $D$ 
\begin{equation}\label{D}
D\indices{^{\mu}_{\nu}} = A\indices{^{\alpha}^{\mu}}A\indices{_{\alpha}_{\nu}}.
\end{equation}
$B$ and $D$ have the same four eigenvalues since from the cyclicity of the trace we 
have $\textrm{Tr}(B^n)=\textrm{Tr}(D^n)$ for all integer $n$.
These common eigenvalues determine the singular values of $A$. 

The relation
\begin{equation}
A\indices{_{\beta}^{\mu}} A\indices{^\beta_\alpha}A\indices{_\nu^\alpha}=D\indices{^\mu_\alpha} A\indices{_\nu^\alpha} = A\indices{_\beta^\mu}
B\indices{^\beta_\nu} 
\label{identity}
\end{equation}
shows that $A$ is an intertwining operator\cite{howie} relating the 
eigenspaces of $B$ and $D$. The eigenspaces of $B$ and $D$ are then 
used to bring 
$A$ to its LSVD form.

{\it{Dominant Energy Condition:}} 
The non-negativity of $\rho$ implies that $\textrm{Tr}\rho(\tau_1\otimes\tau_2) \ge 0$, 
where $\tau_1=n^{\mu}\sigma_{\mu}$ and $\tau_2=m^{\mu}\sigma_{\mu}$ are pure 1-qubit states of two subsystems. We conclude that
\begin{equation}	
A\indices{_{\mu\nu}}n^{\mu}m^{\nu}\geq0 
\label{amunu}
\end{equation}
for all lightlike $n^{\mu},m^{\nu}$. 
This implies that the linear transformation  $A\indices{^{\mu}_{\nu}}$ 
maps causal vectors to causal vectors (see Fig. 1). More explicitly,
$A\indices{^\mu_\nu} n^\nu$ is causal if $n^\nu$ is.
This is 
also true of the transpose
of $A$ ($A\indices{_\mu^\nu}n^\mu$ is causal for $n^\mu$ causal) 
and the composite maps $B$ and $D$. This property of mapping the 
light cone into itself is usually demanded of stress energy tensors
in Relativity, where it is called (see Appendix) the Dominant Energy Condition (DEC)\cite{hawkingandellis}.
\begin{figure}[h!]
		\begin{center}
			\includegraphics[width=0.5\textwidth]{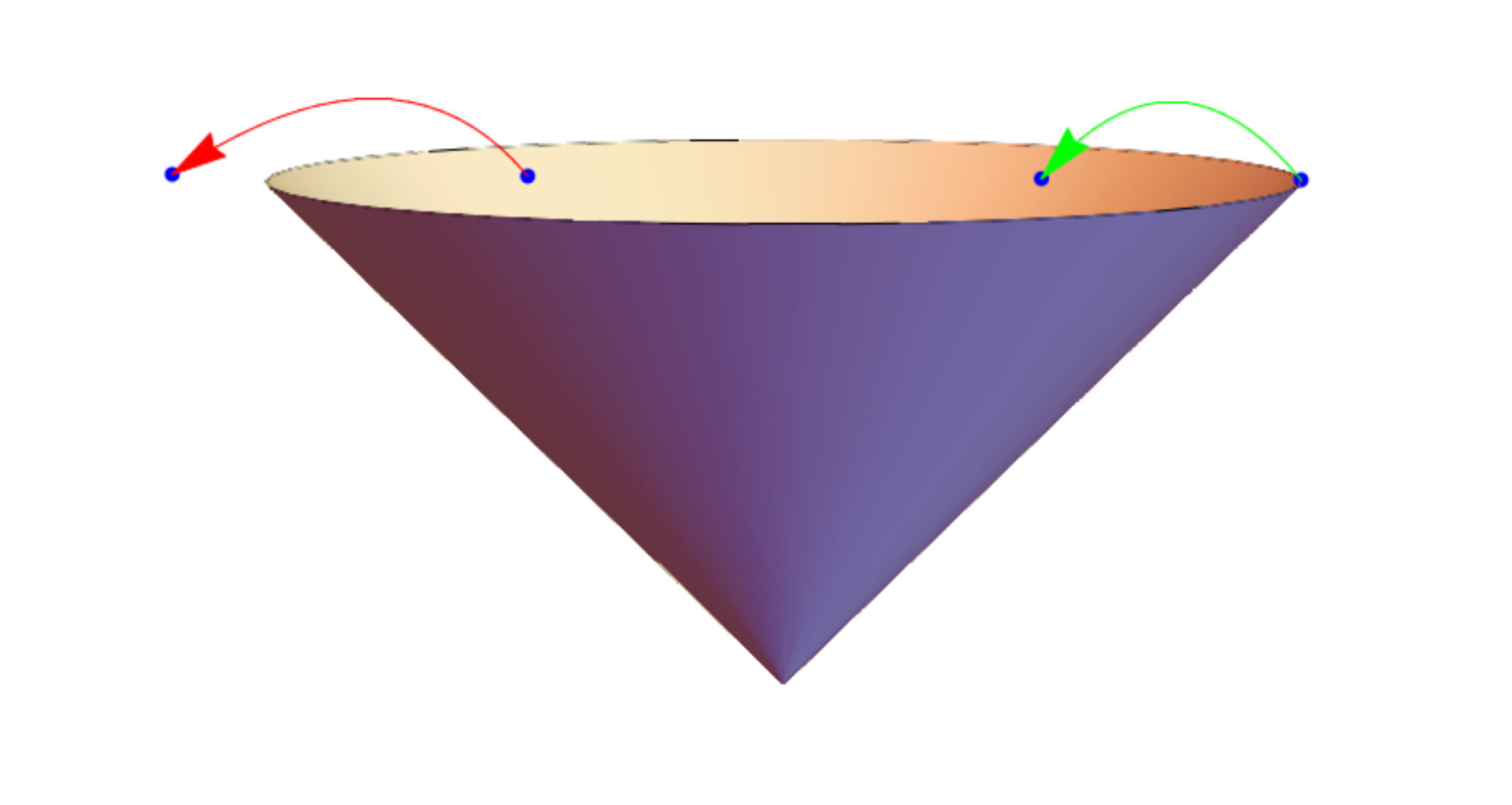}
			\caption{(color online) A representation depicting 
causal vectors getting mapped to causal vectors (green arrow on the right). 
The reverse map of a timelike vector going to a space like 
vector is not allowed (red arrow on the left) by the Dominant Energy Condition.}
		\end{center}
	\end{figure}

%but it can  map a timelike vector
%to the zero vector.  
The dominant energy condition imposes restrictions on the forms that $B$ can take. 
Hawking and Ellis \cite{hawkingandellis} give a classification of the canonical forms 
taken by a symmetric tensor in a Lorentzian space. There are four types, 
of which only Type-I and Type-II are relevant for us, since the others do not
satisfy the DEC. Let $\lambda_0$ be the dominant eigenvalue of $B$ (and $D$).  

{\it Type-I States:} 
These states are defined by the condition that $B$ admits a timelike eigenvector $e_0$ 
(\(B\indices{^{\mu}_{\nu}}e_0^{\nu}=\lambda_0e_0^{\mu} \)) with $\lambda_0>0$.
From Eq. (\ref{identity}) it follows that
$A\indices{_\nu^\alpha} e_0^\nu$ is an eigenvector $E_0^\alpha$ of 
D with the same eigenvalue $\lambda_0$. 
Computing $E_0.E_0=\lambda_0 e_0.e_0$ we see that 
$E_0$ is timelike, since $e_0$ is.
Normalising these eigenvectors, we can write (with $\mu_0>0$),  
\begin{equation}
\mu_0 E_0^\mu=A\indices{^\mu_\alpha} e_0^\alpha
\label{thnik}
\end{equation}

Squaring (\ref{thnik}) we find that  
\begin{equation}
\lambda_0=\mu_0^2.
\label{think}
\end{equation}
Let us define
\[b\indices{^{\mu}_{\nu}}=B\indices{^{\mu}_{\nu}}-\lambda_0
e_0^\mu e^0_\nu\]
$b$ is symmetric and spatial ($b\indices{_{\mu\nu}}=b\indices{_{\nu\mu}},\ b\indices{_{\mu\nu}}e\indices{_0^\nu}=0$) and can therefore be diagonalized by an 
$SO(3)$ transformation. We thus have a diagonal form for $B$.

The orthonormal frame which diagonalises $B$, ($e_a^\mu$) gives us a Lorentz tetrad, whose inverse is $e^a_\mu$. In this frame $B$ has the form:
\begin{equation}\label{Bmn}
B\indices{^{\mu}_{\nu}}= e_a^\mu B\indices{^{a}_{b}} e_\nu^b.
\end{equation}
where $B=diag(\lambda_0,\lambda_1,\lambda_2,\lambda_3)$. 
Similarly
\begin{equation}\label{Dmn}
D\indices{^{\mu}_{\nu}}= E_a^\mu D\indices{^{a}_{b}} E_\nu^b.
\end{equation}
$D=diag(\lambda_0,\lambda_1,\lambda_2,\lambda_3)$. Applying $A$ to $e_a^\nu$ we have
\begin{equation}\label{Amn}
A\indices{^{\mu}_{\nu}}e_a^\nu=\mu_a\delta_a^bE_b^{\mu}=T\indices{_a^b}E_b^\mu
\end{equation}
or equivalently
\begin{equation}\label{Amnn}
A\indices{^{\mu}_{\nu}}=E_b^\mu T\indices{^b_a} e_\nu^a,
\end{equation}

where
$T\indices{^a_b}$ is diagonal with the form
\begin{equation}
T\indices{^a_b}=
\begin{pmatrix}
\mu_0 & 0 & 0 & 0\\
0 & \mu_1 & 0 & 0\\
0 & 0 & \mu_2 & 0\\
0 & 0 & 0 & \mu_3
\label{pmatrixdiag}
\end{pmatrix}
\end{equation}
The $\mu$ s  are the singular values of $A$ 
and $e\indices{_a^{\mu}}$ and $E\indices{^b_{\nu}}$ the 
left and right Partial Lorentz Transformations that bring A to the 
LSVD (Lorentzian Singular Value Decomposition) form (\ref{pmatrixdiag}).
Since the eigenvalues of $B$ are the squares of the singular values of
$A$, it follows that $\lambda$s are positive. At this stage
$\mu_1,\mu_2,\mu_3$ can all have either sign.
By Partial Lorentz transformations 
(e.g by rotation by $\pi$ in the $x-y$ plane) 
it is possible to reverse the signs of two of $\mu_1,\mu_2,\mu_3$. 
By such transformations it is possible to arrange for all 
of $\mu_1,\mu_2,\mu_3$  to have the same sign. 
$\mu_0$, of course, is positive (\ref{think}).

\section{IV States and Separability}
The DEC is a necessary condition for $\rho$ to be a state (have non negative
eigenvalues). From the LSVD form (\ref{pmatrixdiag}) it is easy to write down 
sufficient conditions on the $\mu$s to ensure that $\rho$ is positive.
The diagonal form (\ref{pmatrixdiag}) leads to a state $\rho$ 
\begin{equation}
\left(
\begin{array}{cccc}
\mu_0-\text{$\mu_3$} & 0 & 0 & \text{$\mu_2$}-\text{$\mu_1$} \\
 0 & \text{$\mu_3$}+\mu_0 & -\text{$\mu_1$}-\text{$\mu_2$} & 0 \\
 0 & -\text{$\mu_1$}-\text{$\mu_2$} & \text{$\mu_3$}+\mu_0 & 0 \\
 \text{$\mu_2$}-\text{$\mu_1$} & 0 & 0 & \mu_0-\text{$\mu_3$} \\
\end{array}
\right)
\label{X}
\end{equation}
with eigenvalues 
\begin{eqnarray}\label{mu}
\mu_1 -\mu_2 -\mu_3 &+& \mu_0\nonumber\\
-\mu_1 +\mu_2 -\mu_3 &+& \mu_0\nonumber\\
-\mu_1 -\mu_2 +\mu_3 &+& \mu_0\nonumber\\
\mu_1 +\mu_2 +\mu_3 &+& \mu_0\nonumber\\
\label{eigenvalues}
\end{eqnarray} 
Requiring that the eigenvalues of $\rho$ are positive gives us the 
conditions

\begin{eqnarray}
-\mu_1 +\mu_2 +\mu_3 &\le& \mu_0\nonumber\\
\mu_1 -\mu_2 +\mu_3 &\le& \mu_0\nonumber\\
\mu_1 +\mu_2 -\mu_3 &\le& \mu_0\nonumber\\
\mu_1 +\mu_2 +\mu_3 &\ge& -\mu_0\nonumber\\
\label{stateconditions}
\end{eqnarray}

The form of $T\indices{^a_b}$ gives us a way to express it in separable form, 
provided $T\indices{^a_b}$ (See also the appendix below) satisfies the strong energy condition \cite{hawkingandellis}: 
\[\mu_1+\mu_2+\mu_3\leq \mu_0. \]
Let us define an orthonormal frame $T^a, X^a, Y^a, Z^a$ in which $T\indices{^{a}_{b}}$ is diagonal.
Suppose first that $\mu_1,\mu_2,\mu_3$ are all non negative.
\begin{equation}
T\indices{^{a}_{b}}=\mu_1X^aX_{b}
+\mu_2Y^aY_b
+\mu_3Z^aZ_{b}+\mu_0T^aT_b
\end{equation}
Let us also define lightlike vectors $X_{\pm}=(T\pm X)/\sqrt{2}$ and similarly $Y_{\pm}$ and $Z_{\pm}$. From the identity
\begin{equation}
X_+^aX_{+b}+X_-^aX_{-b}=X^aX_b+T^aT_b 
\label{identity1}
\end{equation}
we can write $T\indices{^{a}_{b}}$ as
\begin{eqnarray}
T\indices{^{a}_{b}}=&&\mu_1(X_+^aX_{+b}+X_-^aX_{-b})\nonumber\\
&+&\mu_2(Y_+^aY_{+b}+Y_-^aY_{-b})\nonumber\\
&+&\mu_3(Z_+^aZ_{+b}+Z_-^aZ_{-b})\nonumber\\
&+&(\mu_0-\mu_1-\mu_2-\mu_3)T^aT_b
\end{eqnarray}
$T$ is explicitly in separable form provided
\[\mu_0 \geq \mu_1+\mu_2+\mu_3,\]
i.e. the Strong Energy Condition(SEC) is satisfied.

If $\mu_1,\mu_2,\mu_3$ are all non positive, they 
automatically satisfy (\ref{stateconditions}) 
$|\mu_1|+|\mu_2|+|\mu_3|\le \mu_0$. 
The identity
\begin{equation}
X_+^aX_{-b}+X_-^aX_{+b}=-X^aX_b+T^aT_b 
\label{identity2}
\end{equation}
gives us 
\begin{eqnarray}
T\indices{^{a}_{b}}=&&|\mu_1|(X_+^aX_{-b}+X_-^aX_{+b})\nonumber\\
&+&|\mu_2|(Y_+^aY_{-b}+Y_-^aY_{+b})\nonumber\\
&+&|\mu_3|(Z_+^aZ_{-b}+Z_-^aZ_{+b})\nonumber\\
&+&(\mu_0-|\mu_1|-|\mu_2|-|\mu_3|)T^aT_b,
\end{eqnarray}
which is in separable form.

Conversely, if $A$ represents a separable state, we can write
\[A_{\mu\nu}=\sum_{i} w_i\ n_{\mu}^i m_{\nu}^i \]
where $w_i>0$ are positive weights and $n^i$ and $m^i$ are 
future pointing causal vectors. Without loss of 
generality we can suppose $n,m$ to be 
lightlike (since time-like vectors are convex combinations of lightlike ones) 
and further absorb $w_i$ into the vectors $n,m$. 
Computing
\begin{eqnarray}
	A_{xx}+A_{yy}+A_{zz}&=&\sum_{i}\vec{n_i}.\vec{m_i}\ 
\leq \sum_{i}|n_i||m_i|\nonumber\\
	&=&\sum_{i}n_{i0}m_{i0}\nonumber\\
	&=&A_{00}	
\end{eqnarray}
Applying this argument to the LSVD diagonal form $T$, we see that 
separable states satisfy the SEC. 
Thus we have shown that the SEC is necessary and 
sufficient for separability. If the SEC is 
satisfied  we find an explicit 
decomposition of $T\indices{^{a}_{b}}$ 
(and therefore of $A$) into separable form.

\section{V. Three dimensional representation of the two-qubit state space}
As we discussed earlier, the 16 dimensional space of un-normalized density matrices 
undergoes a reduction to a 4 parameter family of twelve dimensional fibers under the action of left and right Partial Lorentz 
Transformations. In fact, the 4 parameter $(\mu_0, \mu_1, \mu_2, \mu_3)$ representation can be further 
reduced to a 3 parameter representation since only the ratios 
are relevant. Since we have assumed $\lambda_0\neq0$ we have $\mu_0\neq0$. By scaling let us set $\mu_0=1$ and plot 
a simple three dimensional representation of the state space. 
From the DEC, it follows that $0\le |\mu_{\hat{a}}|\le 1, {\hat{a}}=1,2,3$, 
so the states
lie within the cube  of side $2$  whose body diagonal 
connects ${\bf{\tilde{P}}}=\{-1,-1,-1\}$ to ${\bf{P}}=\{1,1,1\}$. 
\begin{figure}[h!]
		\begin{center}
			\includegraphics[width=0.5\textwidth]{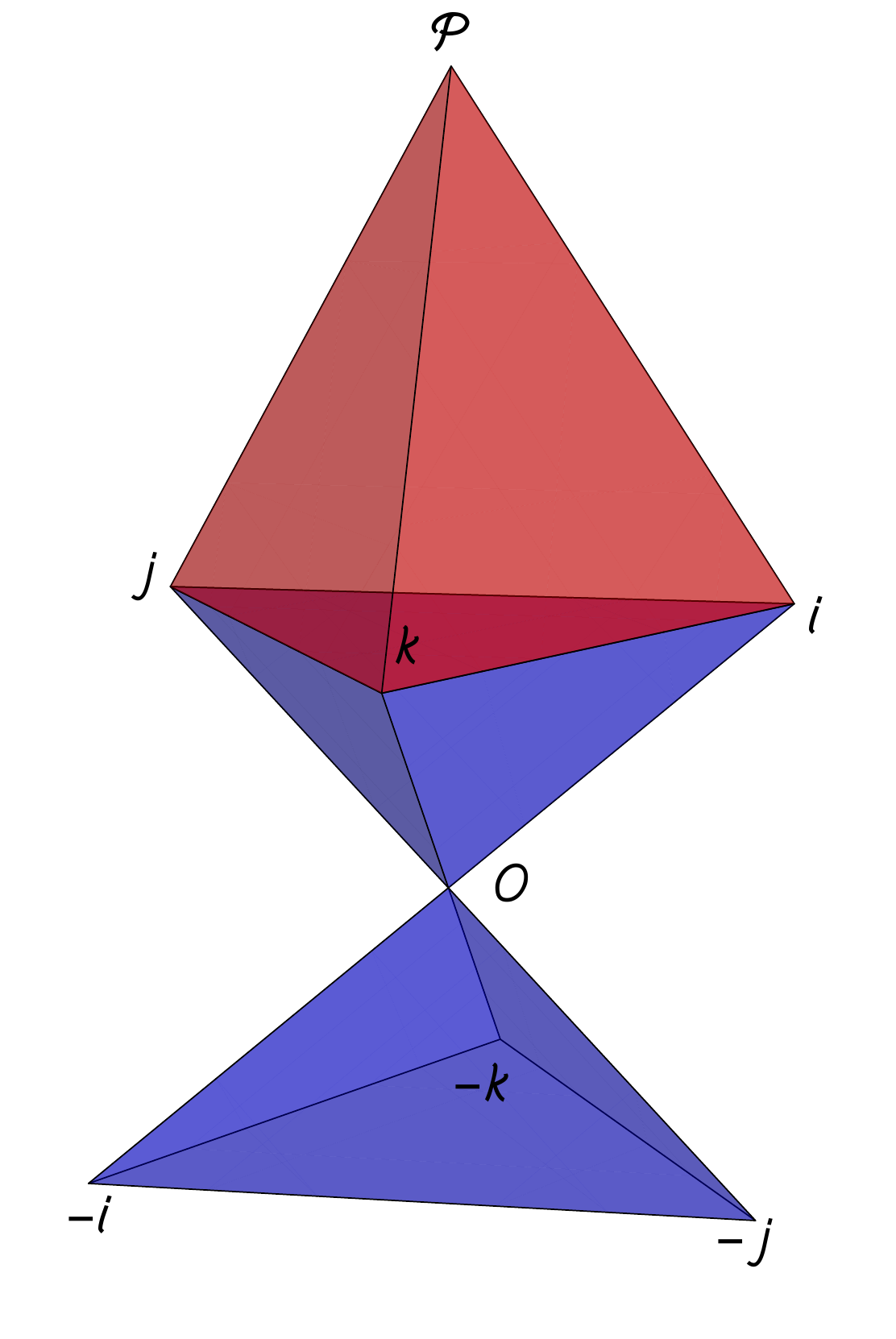}
			\caption{(color online) A 
three dimensional representation of the state space
of $\mu_1,\mu_2,\mu_3$ for Type-I states. 
The red tetrahedron (${\bf \{P,i,j,k}\}$)  
represents the set of entangled states and the blue
tetrahedra (${\bf \{O,i,j,k\}}$ and ${\bf \{O,-i,-j,-k\}}$), 
the set of separable states. 
The boundary between these two sets is defined by a plane passing through
the tips of the unit vectors
${\bf i,j,k}$.} 
		\end{center}
	\end{figure}
As mentioned earlier, we can suppose that $\mu_1,\mu_2,\mu_3$ have the same sign. 
Instead of the eight octants spanned by the cube above, we need
only restrict ourselves to two of the eight octants: the positive octant
and the negative octant. 
This results in the figure shown in Fig. 2. 

The region shaded blue is the set of separable states. 
All states in the negative octant are separable and 
form the convex hull $S^-=H(\bf{O},-\bf{i},-\bf{j},-\bf{k})$ of the origin $\bf{O}$ and  
the tips of the unit vectors $-\bf{i},-\bf{j},-\bf{k}$. 
The plane passing through $-\bf{i},-\bf{j},-\bf{k}$ divides $\mu$s 
satisfying the state conditions(\ref{stateconditions}) from those that don't.
In the positive octant,
the separable states form the convex hull $S^+=H(\bf{O},\bf{i},\bf{j},\bf{k})$ of the origin $\bf{O}$ and  
the tips of the unit vectors $\bf{i},\bf{j},\bf{k}$. 
The plane passing through $\bf{i},\bf{j},\bf{k}$
divides the separable states from the entangled states. All states ``above'' this plane (Fig. 2)
are entangled and shown in red. 

Note also that under inversion, (reversing the sign of all of $\mu_1,\mu_2,\mu_3$), the separable states $S^+$ and $S^-$ exchange
places, but the entangled states are mapped to regions outside the state space. 
In fact, inversion ${\cal I}$ in the $\vec{\mu}$ space is identical to the partial transpose 
(and to the partial inversion). As expected from the PPT test, the entangled states (in red in Fig. 2) are mapped 
outside the state space by the partial transpose operation.

Finally we remark that the states on the boundary of $S^+$ and $S^-$, where
one or more $\mu$ s vanishes have to be identified with their images under inversion. 
With this identification, Fig. 2 gives a complete elucidation of the generic state space. 
Each point in the state space of Fig. 2 represents an equivalence class of states,
all of which are related by partial Lorentz transformations.

The generic state space includes most of the states of the two qubit system, including all
strictly positive density matrices. The non generic states are characterised by the absence of
a timelike eigenvector for $B$ ($D$). We deal with these in the next section titled exceptional states\cite{avron}.

\section{VI Exceptional States}
There are some states which do not admit a timelike eigenvector for $B$ ($D$). 
For this to happen, the dominant eigenvalue $\lambda_0$ has to be degenerate.

\textbf{Type-II States:}\\
These states are characterised
by the fact that $B$ ($D$) has a repeated lightlike 
eigenvector with positive eigenvalue. The dominant eigenvector can be chosen to be $=X_+$. 
For Type-II states, the LSVD matrix $T\indices{^a_b}$ 
is not diagonal but only in Jordan form. The basis which achieves this
form is not a standard Lorentz frame 
$\{T,X,Y,Z\}$ but a null frame $\{X_+,X_-,Y,Z\}$.
The Jordan form is
\begin{equation}
T\indices{^a_b}=
\begin{pmatrix}
\mu_0 & 0 & 0 & 0\\
x & \mu_0 & 0 & 0\\
0 & 0 & \mu_2 & 0\\
0 & 0 & 0 & \mu_3
\label{pmatrixjordan}
\end{pmatrix}
\end{equation}
where $x>0$. (DEC guarantees $x\ge0$, but if $x$ vanishes, $A$ is of Type-I, 
since $B$ has {\it two} distinct lightlike eigenvectors $X_+,X_-$.) 
We have arbitrarily selected $\mu_1$ 
degenerate with $\mu_0$. Since $\mu_1=\mu_0$ is positive, we can 
arrange for $\mu_2,\mu_3$ also to be positive and we have  
\begin{eqnarray}
T\indices{^{a}_{b}}=&&\mu_0(X_-^aX_{+b}+X_+^aX_{-b})\nonumber\\
&+&\mu_2(Y_+^aY_{+b}+Y_-^aY_{-b})\nonumber\\
&+&\mu_3(Z_+^aZ_{+b}+Z_-^aZ_{-b})\nonumber\\
&+&xX_+^aX_{+b}
\end{eqnarray}
The condition that $A$ is defined 
from a state (Eq. (\ref{stateconditions})) requires 
$\mu_2=\mu_3$. From the argument at the end of section IV, we see that these
states are entangled if $\mu_2=\mu_3>0$. 

If $\mu_2=\mu_3=0$, then
\begin{eqnarray}
T\indices{^{a}_{b}}=&&\mu_0(X_-^aX_{+b}+X_+^aX_{-b})+x X_+^a X_{+b}.\nonumber\\
\end{eqnarray}
 These states are clearly in separable form.
The Type-II states are shown in Fig. 3. The blue dots represent the
separable states and the red lines the entangled ones. 
By switching the roles of $B$ and $D$, we also have states where
the Jordan form is the transpose of (\ref{pmatrixjordan}).

\textbf{Type-II0 States:}\\
Finally, we address the possibility that the dominant eigenvalue $\lambda_0$ vanishes. 
As described in the appendix, these states come in three families 
($t$ is a timelike vector and $x$ is positive):
\begin{enumerate}
\item{Type-II0a}: 
$A\indices{^\mu_\nu}=x t\indices{^\mu} l\indices{_\nu}$. $B$ vanishes identically.
\item{Type-II0b}: $A\indices{^\mu_\nu}=x l\indices{^\mu} t\indices{_\nu}$. $D$ vanishes identically.
\item{Type-II0c}: 
$A\indices{^\mu_\nu}=x l\indices{_1^\mu} l\indices{_2_\nu}$. Both $B$ and $D$ vanish.
\end{enumerate}

These states are separable 
and because they have vanishing $\mu_0$, do not find a 
place in either Fig.2 or Fig.3. 
The form of the stress tensor for Type-II0c is 
$T\indices{^a_b}=x X\indices{_+^a} X\indices{_{+b}}$. 
Such a form for the stress tensor appears in Relativity where it is known as a 
null fluid or null dust\cite{hawkingandellis}. It represents radiation 
which is all travelling in the same direction.

To summarise our classification (which is explained in more detail in the appendix),
\begin{enumerate}
\item{Type-I}: $\lambda_0>0$ and $B$ (and $D$) admit a timelike eigenvector.
\item{Type-II}: $\lambda_0>0$ and $B$ (and $D$) has a repeated lightlike eigenvector.
\item{Type-II0}: $\lambda_0=0$. $B$ or $D$ (or both) vanish.

\end{enumerate}

\begin{figure}[h!]
		\begin{center}
			\includegraphics[width=0.5\textwidth]{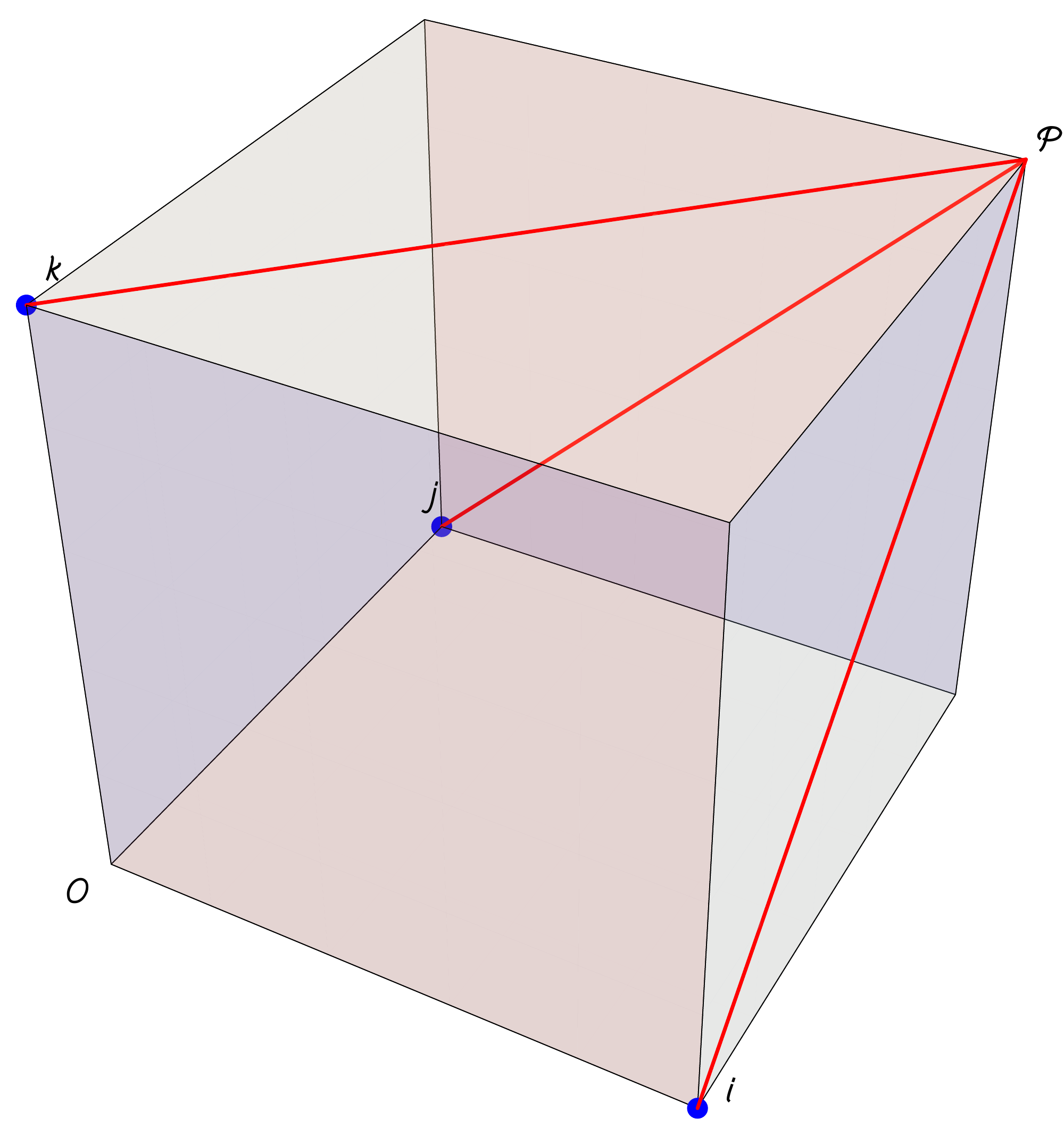}
			\caption{(color online) A three dimensional representation of the state space
for Type-II states.The three blue dots at $\{{\bf i,j,k}\}$ 
represent separable states and the three red lines $\{{\bf i,P } \}$,
$\{{\bf j,P } \}$, $\{{\bf k,P } \}$ represent
entangled states.}
		\end{center}
	\end{figure}

%-------------------------------------------------------------
\section{VII. Conclusion}
We have presented a necessary and sufficient criterion to detect two qubit entanglement.
In addition our approach reveals a separable form of the density matrix if it exists. Our approach is based
on Lorentzian geometry, in particular a Lorentzian Singular Value Decomposition. The LSVD has also been described
by Avron et al \cite{avron}. They also notice 
the relevance of the Dominant Energy Condition that all states must satisfy
and go on to 
give a three dimensional graphical representation of the state space.
However, Avron et al \cite{avron} do not propose an entanglement test, as we have done. 
Neither do they comment on the relevance of the strong energy condition to entanglement.
Our graphical representation,
though related to \cite{avron}, is simpler, because we reduce the picture from 
eight octants to two. 
There has also been work \cite{rudolf}
which proposes an entanglement test based on a standard Singular Value Decomposition. 
However, this test only works on a restricted class of 
states: the reduced density matrices of each subsystem have to be maximally disordered.
We go beyond earlier work in providing an explicit construction of a 
separable state for 
the density matrix in those cases where it exists.

Our focus in this paper is entirely on quantum entanglement. 
There are other quantum correlations like
discord  described for example in \cite{PhysRevA.81.042105}, which are
not considered here. Ref.\cite{PhysRevA.81.042105} 
studies the so-called X states, which have nonzero entries
on the diagonal and the anti-diagonal. The focus of Ref.\cite{PhysRevA.81.042105} is the study of
quantum discord for two qubit X states, with a view to understanding the relation between 
quantum discord, classical correlations and entanglement. They observe that these are independent 
measures of correlation.

Ref. \cite{simon} also addresses X states and quantum discord.
Just as we do here, Ref. \cite{simon} also makes use of Lorentzian structures.
However, the local operations considered are local unitary transformations (six parameters in all)
and the canonical forms used are X states, which are characterised by essentially five parameters. 
As a result the total dimension of the state space explored is generically eleven, which falls short of the dimension
of fifteen, for normalised states. In contrast, our use of local (or partial) Lorentz transformations 
provides twelve parameters, which along with the four eigenvalues of the canonical diagonal form provides a complete
characterisation of the sixteen dimensional unnormalised state space. It is interesting to note that our Eq. (\ref{X}) represents
an X state, but the number of parameters appearing is only four. In our treatment, not all X states are required to 
produce the general state by local Lorentz transformations.

There appears to be a rich Lorentzian structure hidden within the theory of quantum entanglement. 
The relation is probably best appreciated
using spinors, which have been studied by relativists like Penrose, Newman 
etc\cite{penroserindler}. 
In this exposition, we have deliberately avoided
the use of spinor language since this is not widely used in the general physics community. 
The key property of Partial Lorentz Transformations used here
is that they map states to states, separable states to separable states and entangled states to entangled states.
This allows us to decompose the total set of states into equivalence classes. Any two elements from the same equivalence
class are related by Partial Lorentz Transformations and are either both separable or both entangled. To decide
whether a particular equivalence class is entangled or separable, we can choose any element from the class. 
By choosing the canonical form given by the LSVD decomposition, we are able to easily determine if the class is
separable or entangled.

Although the test proposed in \cite{earlier} relies only on the eigenvalues of $B$ ($D$), it is important to realise
that the {\it state} depends both on the eigenvalues and the eigenvectors of $B$ ($D$). While a knowledge of the eigenvalues
is enough to determine if a state is separable, one needs also the eigenvectors to explicitly write out the separable form.

By setting quantum states in correspondence with tensors in Minkowski space, we were naturally led to a 
formalism combining Quantum Information Theory with Relativity. 
While the analogy at this level is a purely formal one, 
it may contain the seeds of some future amalgamation of 
Relativity with Quantum Information Theory. For instance one can consider physical realisations of PLT s by
forming two qubits in an entangled state, separating the qubits and acclerating 
one of them adiabatically to a new Lorentz frame. One would expect the states to transform according to the 
formulae of this paper.

How does this theory work in higher dimensional quantum systems? It would appear that one has to find a maximal group of
transformations which takes states to states and separable states to separable states. These would be the appropriate 
generalisation of PLT s to the higher dimensional case.
Once such a group of entanglement preserving 
transformations is identified
the dimensionality of the problem can be drastically reduced. We hope to interest the quantum information community
in this new approach to the problem of detecting quantum entanglement.

\section{Acknowledgements}
We thank Anirudh Reddy and Rafael Sorkin for discussions.

%\appendix{Energy Conditions}

\section{Appendix A: Energy Conditions}

In this appendix we discuss the energy conditions that come into play in our analysis. 
Given a stress energy tensor $T_{a b}$ one requires it to satisfy
some ``reasonable'' positivity conditions. If $T\indices{^a_b}$ has a timelike eigenvector, 
it can be diagonalised (\cite{hawkingandellis}) and brought to 
the form $T\indices{^a_b}=diag(\epsilon,-p_1,-p_2,-p_3)=diag(\mu_0,\mu_1,\mu_2,\mu_3)$,
where $\epsilon$ is the energy density of matter and $p_1,p_2,p_3$ the principal pressures of the matter fluid. Note that in our context,
the pressures are negative when the $\mu$s are positive. The exceptional case, 
where $T$ has  a repeated lightlike eigenvector represents a null
fluid and this corresponds to the Type-II density matrices mentioned above. Below is a short primer on energy conditions, giving the formal 
definition and a physical interpretation. 
Below we will suppose for illustration that $T$ is Type-I and 
can be diagonalised, which is the generic and most interesting case.
\begin{equation}
	T\indices{_{a b}}=
	\begin{pmatrix}
	\epsilon & 0 & 0 & 0\\
	0 & p_1 & 0 & 0\\
	0 & 0 & p_2 & 0\\
	0 & 0 & 0 & p_3
	\end{pmatrix} 
\end{equation}

\subsection{Weak Energy Condition:}
The weak energy condition (WEC) states that given any 
timelike vector $\xi\indices{^{a}}$ $T$ must satisfy:
\begin{equation}\label{wec}
	T\indices{_{a b}}\xi^{a}\xi^{b}\geq 0
\end{equation}	
This yields

\begin{equation}
	\epsilon +p_{\hat{a}}\geq 0 \qquad \textrm{for} \ \hat{a} =1,2,3
\end{equation}	
The weak energy condition physically represents the idea that all observers must see a postive energy density.
There is no negative mass!
	
\subsection{Dominant Energy Condition:}
The Dominant Energy Condition(DEC) states that:	
given any two lightlike vectors $\xi_{1}^{a}$ and $\xi_{2}^{b}$
\begin{equation}
	T\indices{_{a b}}\xi_{1}^{a}\xi_{2}^{b} \geq 0
\label{dec}
\end{equation}
Notice that for $\xi_{1}=\xi_{2}$ we recover the weak energy condition. 
So, the DEC implies the WEC. 
It is enough to demand 
(\ref{dec}) for lightlike $n,m$. 
Since timelike vectors are convex combinations of lightlike ones, it follows that (\ref{dec}) holds for 
timelike $n,m$.
For a suitable choice of $\xi_{1},\xi_{2}$ the DEC gives us:
$\epsilon\ge |p_{\hat{a}}|$ for $\hat{a}=1,2,3$.
The Dominant energy condition requires that all observers see a non spacelike 
matter current $j_a=T_{a b} \xi^b$.
Matter cannot travel faster than light!

\subsection{Strong Energy Condition:}
The strong energy condition(SEC) reads:
\begin{equation}
(T\indices{_{a b}}-\frac{1}{2}Tg\indices{_{a b}})\xi^{a}\xi^{b}\geq 0,\ \boldsymbol{\forall}\  \textrm{time-like}\ \xi
\end{equation}	
We find that the SEC gives us
$\epsilon+p_{\hat{a}}\ge0$ and $\epsilon+p_1+p_2+p_3\ge0.$
The strong energy condition emerges from the 
focussing property of timelike geodesics with tangent vector $\xi^a$ as described by Raychaudhuri's 
equation\cite{raychaudhuri}. 
The focussing of timelike geodesics is determined by the sign of
$R_{a b}\xi^a\xi^b$, where $R_{a b}$ is the Ricci tensor. 
The positivity of $R_{a b}\xi^a\xi^b$ is essentially the SEC via
Einstein's equations.
These ``Energy conditions'' are imposed in Relativity as ``reasonable''. 
They are obeyed by the known classical forms of matter.
However, they are violated by quantum matter and 
Dark Energy violates the SEC. The point $\bf{P}$ in Fig.2 
has a stress energy tensor
of the same form as Dark Energy.

%-------------------------------------------------------------	
\section{Appendix B: Classification of States}
In the text, the division of states into different types is only briefly
described with a reference to Hawking and Ellis \cite{hawkingandellis}. 
Ref.\cite{hawkingandellis} gives four possible types for the stress tensor.
Of these, Type-III and Type-IV violate the weak energy condition and therefore
also the dominant energy condition. These types are irrelevant to 
our present context, since {\it all} states satisfy the DEC. Here we describe
briefly our classification of states into Type-II0, Type-I and Type-II. 
Our Type-II0 is contained in Hawking's Type-II. We separate it
from Type-II because it does not fit into the graphical representation for Type-II states.

To classify the states,
we look at the action of $A\indices{^\mu_\nu}$
on lightlike vectors. Are there lightlike vectors which are mapped to the zero 
vector? If the answer is yes, the state is \\
{\bf Type-II0:} This is further divided
into three classes as follows.\\
Type-II0a: $A$ takes  some lightlike vector $l^\nu$ to zero. 
$A\indices{^\mu_\nu} 
l^\nu=0$. Contracting with an arbitrary timelike 
covector $\alpha_\mu$, and noting that $\alpha_\mu A\indices{^\mu_\nu}$ is
causal and orthogonal to $l^\nu$ 
we see that $A$ must take the form
\begin{equation}
A\indices{^\mu_\nu}=x t^\mu l_\nu
\label{type0a}
\end{equation}
where $x$ is positive, $t$ timelike and $l,t$ normalised by $t.t=l.t=1$. 
This form is Type-II0a. In this case $B$ vanishes and $D\indices{^\mu_\nu}= x^2 l^\mu l_\nu$.

{\bf Type-II0b:} 
The transpose of $A$  takes  some lightlike vector $l^\nu$ to zero. 
$A\indices{_\mu^\nu} 
l^\mu=0$. Contracting with an arbitrary timelike 
covector $\alpha_\nu$, and noting that $\alpha_\nu A\indices{_\mu^\nu}$ is
causal and orthogonal to $l^\mu$ 
we see that $A$ must take the form
\begin{equation}
A\indices{_\mu^\nu}=x l_\mu t^\nu 
\label{type0b}
\end{equation}
where $x$ is positive, $t$ timelike and $l,t$ normalised by $t.t=l.t=1$. 
In this case $D$ vanishes and $B\indices{^\mu_\nu}= x^2 l^\mu l_\nu$.

{\bf Type-II0c:} Both $A$ and the  transpose of $A$  takes  some lightlike vector to zero. 
$A\indices{^\mu_\nu} 
l_1^\nu=0$ and $A\indices{_\mu^\nu} 
l_2^\mu=0$. 
Arguing similarly, 
we see that $A$ must take the form
\begin{equation}
A\indices{^\mu_\nu}=x l_2^\mu l_{1\nu} 
\label{type0c}
\end{equation}
where $x$ is positive, $l_1$ and $l_2$ lightlike and $l_1,l_2$ normalised by $l_1.l_2=1$. 
This form is Type-II0c. In this case both $B$ and $D$ vanish. 
 
If no lightlike vectors are mapped to zero by $A$ or its transpose, we ask how many lightlike
vectors mapped by $A$ (or its transpose) to lightlike vectors.  
If the answer is exactly one, the state
is of  \\
{\bf Type-II:}
We have 
\begin{equation}
A\indices{^\mu_\nu} l^\nu=\mu_0n^\mu
\label{lltoll}
\end{equation}
with $\mu_0>0$.
It follows that the transpose of $A$ maps $n$ to $l$	
\begin{equation}
A\indices{_\nu^\mu} n^\nu=\mu_0 l^\mu
\label{lltolltrans}
\end{equation}
and that $D$ and $B$ have a single lightlike eigenvector 
\begin{equation}
D\indices{^\mu_\nu} l^\nu=\mu_0^2 l^\mu\\
\label{lltolld}
\end{equation}
\begin{equation}
B\indices{^\mu_\nu} n^\nu=\mu_0^2 n^\mu
\label{lltollb}
\end{equation}
In this case $B$ and $D$ can only be brought to Jordan form (\ref{pmatrixjordan}).

{\bf Type-I}
If $A$ maps two (or more) distinct lightlike vectors $l_1^\mu$ and $l_2^\mu$ to 
lightlike vectors $n_1^\mu$ and $n_2^\mu$, the same argument shows that $B$ has 
{\it two} (or more) {\it distinct} lightlike eigenvectors with the same eigenvalue.
If $B$ ($D$) has two 
{\it distinct} lightlike eigenvectors $X_+$ and $X_-$ 
with the same eigenvalue $\lambda_0$,
$B$ also admits a timelike eigenvector $X_-+X_+$ and thus is Type-I.

If there are no lightlike vectors mapped to lightlike vectors by $A$,
$A\indices{^\mu_\nu}l^\nu$ is strictly timelike for all lightlike $l$. We have
a strict version of the DEC.
\begin{equation}
l^\mu A\indices{_\mu_\nu}n^\nu>0
\label{sdec}
\end{equation} 
This implies that $A$, its transpose and the composites $B$ and $D$ map
lightlike vectors to timelike vectors.
To classify the remaining states, let us consider the function
$f(l,n)$ defined on the space of distinct lightlike directions determined by the lightlike
vectors $l$ and $n$. ($l.l=n.n=0$)
\begin{equation}
f(l,n):= \frac{B\indices{_\mu_\nu} l^\mu n^\nu}{l.n}
\label{function}
\end{equation}
By construction $f(l,n)$ depends only on the lightlike directions of $l,n$. 
By (\ref{sdec}), the 
numerator is positive and the function $f(l,n)$ approaches positive
infinity as $l$ approaches $n$. The global minimum of $f$ occurs 
at $l_0,n_0$ with $l_0$ and $n_0$ linearly independent lightlike vectors, which
we can normalise by $l_0.n_0=1$. 
By considering the first
variation of $f$ around its minimum, we see that the $l_0,n_0$ plane
is mapped to itself by $B$:
\begin{eqnarray}
Bl_0&=&\alpha l_0+\beta n_0\\
Bn_0&=&\gamma l_0+\alpha n_0,
\label{Bform}
\end{eqnarray}
where $\alpha=B(l_0,n_0),
\beta=B(l_0,l_0),\gamma=B(n_0,n_0)$ are all strictly positive 
by (\ref{sdec}). 
It is easily seen that $B$ has dominant eigenvalue 
$\lambda_0=\alpha+\sqrt{\beta \gamma}$
and dominant eigenvector $l_0+(\sqrt{\beta/\gamma})n_0$,
whose norm $2 \sqrt{\beta/\gamma}$ is strictly positive. 
The dominant eigenvector is timelike and the 
state is Type-I. 
This is in fact the generic case and most of the states
of the two qubit system fall in this category. In fact,
 all the interior states where the eigenvalues of $\rho$ 
are strictly positive
fall into Type-I.

%\bibliography{LSVD}

\begin{thebibliography}{12}
\expandafter\ifx\csname natexlab\endcsname\relax\def\natexlab#1{#1}\fi
\expandafter\ifx\csname bibnamefont\endcsname\relax
  \def\bibnamefont#1{#1}\fi
\expandafter\ifx\csname bibfnamefont\endcsname\relax
  \def\bibfnamefont#1{#1}\fi
\expandafter\ifx\csname url\endcsname\relax
  \def\url#1{\texttt{#1}}\fi
\expandafter\ifx\csname urlprefix\endcsname\relax\def\urlprefix{URL }\fi
\providecommand*{\bibinfo}[2]{#2}
\providecommand*{\eprint}[1]{#1}
\providecommand*{\url}[1]{#1}
\begingroup\makeatletter
 \@temptokena{%
  \expandafter\ifx\csname citenamefont\endcsname\relax
   \DeclareRobustCommand\citenamefont{\@firstofone}%
   \global\let\citenamefont\citenamefont
   \global\expandafter\let\csname citenamefont \expandafter\endcsname\csname
  citenamefont \endcsname
  \fi
 }\if@filesw\immediate\write\@auxout{\the\@temptokena}\fi
\expandafter\endgroup\the\@temptokena

\bibitem[{\citenamefont{Peres}(1996)}]{PhysRevLett.77.1413}
\bibinfo{author}{\bibfnamefont{A.}~\bibnamefont{Peres}},
  \bibinfo{journal}{Phys. Rev. Lett.} \textbf{\bibinfo{volume}{77}},
  \bibinfo{pages}{1413} (\bibinfo{year}{1996}),
  \urlprefix\url{https://link.aps.org/doi/10.1103/PhysRevLett.77.1413}.

\bibitem[{\citenamefont{Horodecki} \emph{et~al.}(1996)\citenamefont{Horodecki,
  Horodecki, and Horodecki}}]{horodeckipla}
\bibinfo{author}{\bibfnamefont{M.}~\bibnamefont{Horodecki}},
  \bibinfo{author}{\bibfnamefont{P.}~\bibnamefont{Horodecki}},
  \bibnamefont{and}
  \bibinfo{author}{\bibfnamefont{R.}~\bibnamefont{Horodecki}},
  \bibinfo{journal}{Physics Letters A}
  \textbf{\bibinfo{volume}{223}}(\bibinfo{number}{1}), \bibinfo{pages}{1 }
  (\bibinfo{year}{1996}), ISSN \bibinfo{issn}{0375-9601},
  \urlprefix\url{http://www.sciencedirect.com/science/article/pii/S0375960196007062}.

\bibitem[{\citenamefont{Bengtsson and Zyczkowski}(2007)}]{geometry}
\bibinfo{author}{\bibfnamefont{I.}~\bibnamefont{Bengtsson}} \bibnamefont{and}
  \bibinfo{author}{\bibfnamefont{K.}~\bibnamefont{Zyczkowski}},
  \emph{\bibinfo{title}{Geometry of Quantum States: An Introduction to Quantum
  Entanglement}} (\bibinfo{publisher}{Cambridge University Press},
  \bibinfo{year}{2007}), ISBN \bibinfo{isbn}{9781139453462},
  \urlprefix\url{https://books.google.co.in/books?id=aA4vXMbuOTUC}.

\bibitem[{\citenamefont{Samuel} \emph{et~al.}(2017)\citenamefont{Samuel,
  Shivam, and Sinha}}]{earlier}
\bibinfo{author}{\bibfnamefont{J.}~\bibnamefont{Samuel}},
  \bibinfo{author}{\bibfnamefont{K.}~\bibnamefont{Shivam}}, \bibnamefont{and}
  \bibinfo{author}{\bibfnamefont{S.}~\bibnamefont{Sinha}},
  \bibinfo{journal}{https://arxiv.org/abs/1712.06801}  (\bibinfo{year}{2017}).

\bibitem[{\citenamefont{Duplij} \emph{et~al.}(2004)\citenamefont{Duplij,
  Siegel, and Bagger}}]{howie}
\bibinfo{editor}{\bibfnamefont{S.}~\bibnamefont{Duplij}},
  \bibinfo{editor}{\bibfnamefont{W.}~\bibnamefont{Siegel}}, \bibnamefont{and}
  \bibinfo{editor}{\bibfnamefont{J.}~\bibnamefont{Bagger}}, eds.,
  \emph{\bibinfo{title}{Concise Encyclopedia of Supersymmetry: And
  noncommutative structures in mathematics and physics}}
  (\bibinfo{publisher}{Springer Netherlands}, \bibinfo{year}{2004}), pp.
  \bibinfo{pages}{208--208}, ISBN \bibinfo{isbn}{978-1-4020-4522-6},
  \urlprefix\url{https://doi.org/10.1007/1-4020-4522-0_271}.

\bibitem[{\citenamefont{Hawking and Ellis}(1973)}]{hawkingandellis}
\bibinfo{author}{\bibfnamefont{S.}~\bibnamefont{Hawking}} \bibnamefont{and}
  \bibinfo{author}{\bibfnamefont{G.}~\bibnamefont{Ellis}},
  \emph{\bibinfo{title}{The Large Scale Structure of Space-Time}}, Cambridge
  Monographs on Mathematical Physics (\bibinfo{publisher}{Cambridge University
  Press}, \bibinfo{year}{1973}), ISBN \bibinfo{isbn}{9780521099066},
  \urlprefix\url{https://books.google.co.in/books?id=QagG\_KI7Ll8C}.

\bibitem[{\citenamefont{Avron and Kenneth}(2009)}]{avron}
\bibinfo{author}{\bibfnamefont{J.}~\bibnamefont{Avron}} \bibnamefont{and}
  \bibinfo{author}{\bibfnamefont{O.}~\bibnamefont{Kenneth}},
  \bibinfo{journal}{Annals of Physics}
  \textbf{\bibinfo{volume}{324}}(\bibinfo{number}{2}), \bibinfo{pages}{470 }
  (\bibinfo{year}{2009}), ISSN \bibinfo{issn}{0003-4916},
  \urlprefix\url{http://www.sciencedirect.com/science/article/pii/S0003491608001188}.

\bibitem[{\citenamefont{Rudolph}(2003)}]{rudolf}
\bibinfo{author}{\bibfnamefont{O.}~\bibnamefont{Rudolph}},
  \bibinfo{journal}{Phys. Rev. A} \textbf{\bibinfo{volume}{67}},
  \bibinfo{pages}{032312} (\bibinfo{year}{2003}),
  \urlprefix\url{https://link.aps.org/doi/10.1103/PhysRevA.67.032312}.

\bibitem[{\citenamefont{Ali} \emph{et~al.}(2010)\citenamefont{Ali, Rau, and
  Alber}}]{PhysRevA.81.042105}
\bibinfo{author}{\bibfnamefont{M.}~\bibnamefont{Ali}},
  \bibinfo{author}{\bibfnamefont{A.~R.~P.} \bibnamefont{Rau}},
  \bibnamefont{and} \bibinfo{author}{\bibfnamefont{G.}~\bibnamefont{Alber}},
  \bibinfo{journal}{Phys. Rev. A} \textbf{\bibinfo{volume}{81}},
  \bibinfo{pages}{042105} (\bibinfo{year}{2010}),
  \urlprefix\url{https://link.aps.org/doi/10.1103/PhysRevA.81.042105}.

\bibitem[{\citenamefont{Sabapathy and Simon}(2013)}]{simon}
\bibinfo{author}{\bibfnamefont{K.~K.} \bibnamefont{Sabapathy}}
  \bibnamefont{and} \bibinfo{author}{\bibfnamefont{R.}~\bibnamefont{Simon}}
  (\bibinfo{year}{2013}), \urlprefix\url{https://arxiv.org/pdf/1311.0210.pdf}.

\bibitem[{\citenamefont{Penrose and Rindler}(1984)}]{penroserindler}
\bibinfo{author}{\bibfnamefont{R.}~\bibnamefont{Penrose}} \bibnamefont{and}
  \bibinfo{author}{\bibfnamefont{W.}~\bibnamefont{Rindler}},
  \emph{\bibinfo{title}{The geometry of world-vectors and spin-vectors}}
  (\bibinfo{publisher}{Cambridge University Press},
  \bibinfo{year}{1984})\emph{\bibinfo{series}{Cambridge Monographs on
  Mathematical Physics} of vol. \bibinfo{volume}{1}}, p.
  \bibinfo{pages}{1–67}.

\bibitem[{\citenamefont{Raychaudhuri}(1955)}]{raychaudhuri}
\bibinfo{author}{\bibfnamefont{A.}~\bibnamefont{Raychaudhuri}},
  \bibinfo{journal}{Phys. Rev.} \textbf{\bibinfo{volume}{98}},
  \bibinfo{pages}{1123} (\bibinfo{year}{1955}),
  \urlprefix\url{https://link.aps.org/doi/10.1103/PhysRev.98.1123}.

\end{thebibliography}

\end{document}